\documentclass[aps,nofootinbib]{revtex4}
\def\be{\begin{equation}}
\def\ee{\end{equation}}
\def\ba{\begin{eqnarray}}
\def\ea{\end{eqnarray}}

\usepackage{amsmath,bbm}
\usepackage{amssymb}
\usepackage{mathrsfs}
\usepackage{mathtools}
\usepackage{amsthm}
\usepackage{xcolor}
\usepackage{verbatim}
\usepackage[pdftex]{graphicx}
\usepackage{hyperref}
\hypersetup{
	colorlinks=true,
	linkcolor=blue,
	filecolor=magenta,      
	urlcolor=cyan,
}
\newcommand{\R}{\mathbb{R}}

\DeclareFontFamily{U}{rsfs}{}         
\DeclareFontShape{U}{rsfs}{m}{n}{<5> rsfs5 <6><7> rsfs7          %
  <8><9><10><10.95><12><14.4><17.28><20.74><24.88> rsfs10}{}     %
\DeclareMathAlphabet{\mathfs}{U}{rsfs}{m}{n}                     %
\newcommand{\mfs}[1]{\mathfs {#1}}                               %
\newcommand{\n}{{\nonumber}}

\newcommand{\sF}{{\mfs F}}

\newcommand{\sH}{{\mfs H}}

 \usepackage{braket}

\newcommand{\dd}{\mathrm{d}}

\newcommand{\RN}[1]{%
  \textup{\uppercase\expandafter{\romannumeral#1}}%
}

\newcommand{\va}{\scriptscriptstyle}

\usepackage{braket}

\RequirePackage[normalem]{ulem} 
\RequirePackage{color}\definecolor{RED}{rgb}{1,0,0}\definecolor{BLUE}{rgb}{0,0,1} 
\providecommand{\DIFaddbegin}{} 
\providecommand{\DIFaddend}{} 
\providecommand{\DIFdelbegin}{} 
\providecommand{\DIFdelend}{} 
\providecommand{\DIFaddbeginFL}{} 
\providecommand{\DIFaddendFL}{} 
\providecommand{\DIFdelbeginFL}{} 
\providecommand{\DIFdelendFL}{} 
\newcommand{\DIFscaledelfig}{0.5}
\RequirePackage{settobox} 
\RequirePackage{letltxmacro} 
\newsavebox{\DIFdelgraphicsbox} 
\newlength{\DIFdelgraphicswidth} 
\newlength{\DIFdelgraphicsheight} 
\LetLtxMacro{\DIFOincludegraphics}{\includegraphics} 
\newcommand{\DIFaddincludegraphics}[2][]{{\color{blue}\fbox{\DIFOincludegraphics[#1]{#2}}}} 
\newcommand{\DIFdelincludegraphics}[2][]{
\sbox{\DIFdelgraphicsbox}{\DIFOincludegraphics[#1]{#2}}
\settoboxwidth{\DIFdelgraphicswidth}{\DIFdelgraphicsbox} 
\settoboxtotalheight{\DIFdelgraphicsheight}{\DIFdelgraphicsbox} 
\scalebox{\DIFscaledelfig}{
\parbox[b]{\DIFdelgraphicswidth}{\usebox{\DIFdelgraphicsbox}\\[-\baselineskip] \rule{\DIFdelgraphicswidth}{0em}}\llap{\resizebox{\DIFdelgraphicswidth}{\DIFdelgraphicsheight}{
\setlength{\unitlength}{\DIFdelgraphicswidth}
\begin{picture}(1,1)
\thicklines\linethickness{2pt} 
{\color[rgb]{1,0,0}\put(0,0){\framebox(1,1){}}}
{\color[rgb]{1,0,0}\put(0,0){\line( 1,1){1}}}
{\color[rgb]{1,0,0}\put(0,1){\line(1,-1){1}}}
\end{picture}
}\hspace*{3pt}}} 
} 
\LetLtxMacro{\DIFOaddbegin}{\DIFaddbegin} 
\LetLtxMacro{\DIFOaddend}{\DIFaddend} 
\LetLtxMacro{\DIFOdelbegin}{\DIFdelbegin} 
\LetLtxMacro{\DIFOdelend}{\DIFdelend} 
\DeclareRobustCommand{\DIFaddbegin}{\DIFOaddbegin \let\includegraphics\DIFaddincludegraphics} 
\DeclareRobustCommand{\DIFaddend}{\DIFOaddend \let\includegraphics\DIFOincludegraphics} 
\DeclareRobustCommand{\DIFdelbegin}{\DIFOdelbegin \let\includegraphics\DIFdelincludegraphics} 
\DeclareRobustCommand{\DIFdelend}{\DIFOaddend \let\includegraphics\DIFOincludegraphics} 
\LetLtxMacro{\DIFOaddbeginFL}{\DIFaddbeginFL} 
\LetLtxMacro{\DIFOaddendFL}{\DIFaddendFL} 
\LetLtxMacro{\DIFOdelbeginFL}{\DIFdelbeginFL} 
\LetLtxMacro{\DIFOdelendFL}{\DIFdelendFL} 
\DeclareRobustCommand{\DIFaddbeginFL}{\DIFOaddbeginFL \let\includegraphics\DIFaddincludegraphics} 
\DeclareRobustCommand{\DIFaddendFL}{\DIFOaddendFL \let\includegraphics\DIFOincludegraphics} 
\DeclareRobustCommand{\DIFdelbeginFL}{\DIFOdelbeginFL \let\includegraphics\DIFdelincludegraphics} 
\DeclareRobustCommand{\DIFdelendFL}{\DIFOaddendFL \let\includegraphics\DIFOincludegraphics} 
\RequirePackage{listings} 
\RequirePackage{color} 
\lstdefinelanguage{DIFcode}{ 
  moredelim=[il][\color{red}\sout]{\%DIF\ <\ }, 
  moredelim=[il][\color{blue}\uwave]{\%DIF\ >\ } 
} 
\lstdefinestyle{DIFverbatimstyle}{ 
	language=DIFcode, 
	basicstyle=\ttfamily, 
	columns=fullflexible, 
	keepspaces=true 
} 
\lstnewenvironment{DIFverbatim}{\lstset{style=DIFverbatimstyle}}{} 
\lstnewenvironment{DIFverbatim*}{\lstset{style=DIFverbatimstyle,showspaces=true}}{} 

\begin{document}

\title{Light-cone thermodynamics: purification of the Minkowski vacuum}

\author{Alejandro Perez}
\email{perez@cpt.univ-mrs.fr}
\affiliation{Aix Marseille Univ, Universit\'e de Toulon, CNRS, CPT, 13000 Marseille, France}
\author{Salvatore Ribisi}
\email{salvatore.ribisi@cpt.univ-mrs.fr}
\affiliation{Aix Marseille Univ, Universit\'e de Toulon, CNRS, CPT, 13000 Marseille, France}

\begin{abstract}

We explicitly express the Minkowski vacuum of a massless scalar field in terms of the particle notion associated with suitable spherical conformal killing fields. These fields are orthogonal to the light wavefronts originating from a sphere with a radius of $r_{\va H}$ in flat spacetime: a bifurcate conformal killing horizon that exhibits  semiclassical features similar to those of black hole horizons and Cauchy horizons of spherically symmetric black holes. Our result highlights the quantum aspects of this analogy and extends the well-known decomposition of the Minkowski vacuum in terms of Rindler modes, which are associated with the boost Killing field normal to a pair of null planes in Minkowski spacetime (the basis of the Unruh effect). While some features of our result have been established by Kay and Wald's theorems in the 90s---on quantum field theory in stationary spacetimes with bifurcate Killing horizons---the added value we provide here lies in the explicit expression of the vacuum. 
%
\end{abstract}

\maketitle

\section{Introduction}
Light cones emanating from a sphere of radius $r_H$ in Minkowski spacetime satisfy a set of laws which are the analog of the thermodynamic laws satisfied by black holes \cite{DeLorenzo:2017tgx}. When tested with conformally invariantly coupled scalar fields, this is a consequence of the fact that the geometry and (conformal) symmetries of flat spacetime coincides with the geometry and symmetries of certain stationary black hole solutions \cite{DeLorenzo:2018ghq}.  
In addition to satisfying a version of the zeroth, first, second, and third laws of black hole mechanics, it is shown that suitable accelerated observers following the orbits of spherical conformal Killing vector fields (for whom the light cones are conformal Killing horizons) perceive the Minkowski vacuum as a thermal state with a (conformal) temperature $T=\kappa/(2\pi)$ with $\kappa$ a natural notion of surface gravity (we review some of the details below). Even though the thermality of the vacuum was derived explicitly in this work,  it can be seen as the consequence of a simple adaptation to conformal Killing bifurcate horizon of general theorems  by Wald and Kay \cite{Kay:1988mu}. While there is a vast literature devoted to the restriction of the Minkowski vacuum to the domain of dependence of a ball \cite{Hislop:1981uh, Martinetti:2002sz, Martinetti:2008ja, Casini:2011kv, Jacobson:2022gmo} (and generalizations to maximally symmetric spaces \cite{Jacobson:2018ahi}), here we are interested in the description of the Minkowski vacuum in the entire spacetime. In particular, it is the causal complement of the so-called diamond that actually represents (in the sense of  \cite{DeLorenzo:2017tgx}) the region accessible to the (analog of the) outside stationary observers of a Black hole. 
 We will show that the Minkowski vacuum can be decomposed in terms of modes whose time evolution is adapted to conformal Killing vector fields of flat spacetime that have the previously mentioned light cones as horizons. Our result is the generalization of  Unruh's  \cite{Unruh:1976db} where, instead of uniformly accelerated observes moving away from a plane wave that defines their Rindler Horizon, we have a family of suitable radially accelerating observers moving away from a spherical wave front of radius $r_H$ at $t=0$. Instead of trying to compute explicitly Bogoliubov coefficients via inner products and projections, we will use Unruh's original technique consisting on characterizing positive frequency modes by their analyticity properties in the complex plane of complexified time.

\section{The spherical conformal Killing fields of interest}

\begin{figure}[h!!!!!!!!!]
	\centering
	{\includegraphics[scale=0.6]{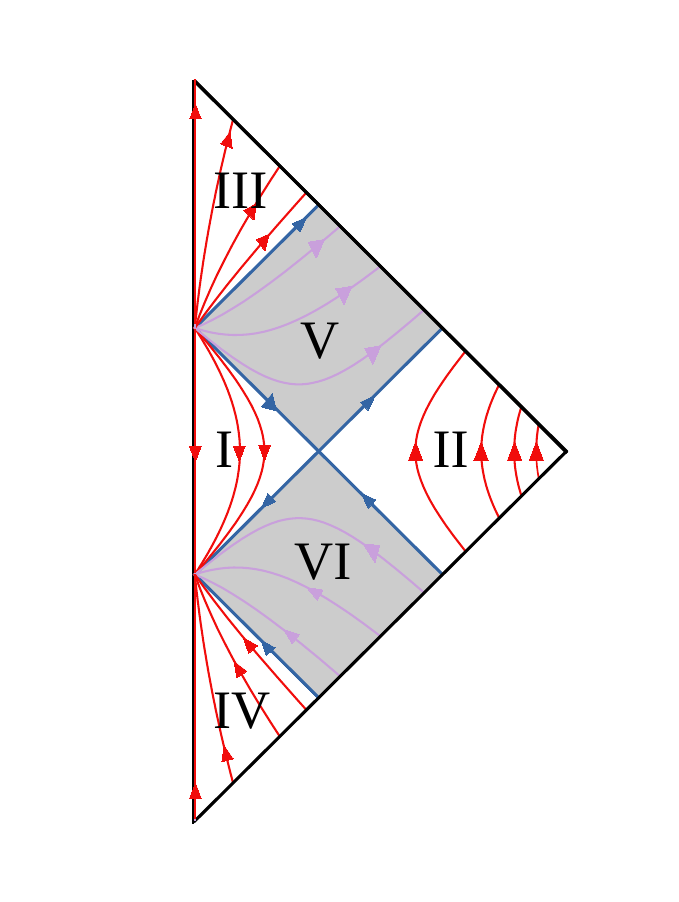}}
	\caption{Integral lines of the spherical conformal Killing field \eqref{kikin}. The vector field becomes null on the light wave-fronts emanating from a sphere of radius $r_H$ (by choice here at $t=0$). These light fronts are bifurcate conformal Killing horizons with bifurcation surface given by the same sphere.}
	\label{figurete}
\end{figure}

We start from the Minkowski metric in spherical coordinates
\begin{align}
	\dd s^2 &= -\dd t^2 + \dd r^2 + r^2 \dd \theta^2 + r^2 \sin^2(\theta) \ \dd   \varphi^2 ,
\end{align}
then we introduce advanced and retarded time null coordinates  
\ba\label{uv} v &\equiv& t + r, \n \\ u &\equiv& t - r.\ea 
In terms of these, the spherical conformal Killing 
fields of our interest are written as 
\begin{align}\label{kikin}
	\xi^a &= \frac{v^2 - r_H^2}{r_0^2 - r_H^2} \left( \frac{\partial}{\partial v} \right)^a + \frac{u^2 - r_H^2}{r_0^2 - r_H^2} \left( \frac{\partial}{\partial u} \right)^a ,
\end{align}
which are completely characterized by two parameters $r_0$ and $r_H$.
Its norm is given by
\begin{align}\label{kiki}
	\xi \cdot \xi &= - \frac{\left( v^2 - r_H^2 \right) \left( u^2 - r_H^2 \right)}{\left( r_0^2 - r_H^2 \right)^2} ,
\end{align}
whose sign divides flat spacetime in six different regions separated by a bifurcate conformal killing horizon 
where it vanishes (at the light fronts $u=\pm r_H$ and $v=\pm r_H$), see Figure \ref{figurete}.
The interpretation of the free parameters is the following: at $t=0$ the sphere of radius $r_0$ is a sphere where $\xi\cdot\xi=-1$, while 
the sphere of radius $r_H$ is the place where the conformal Killing field vanishes (the bifurcating sphere).  
We will assume that $r_0>r_H$, so that the Killing is normalized somewhere in the outside region of the black hole analog in the sense of  \cite{DeLorenzo:2017tgx, DeLorenzo:2018ghq}.  The vector field \eqref{kiki} is null on the light cones defined by
\begin{align}
	u = u_\pm = \pm r_H , \qquad v = v_\pm = \pm r_H .
\end{align}
Hence the conformal Killing vector field  \eqref{kikin}  divides the spacetime in six separate regions (see Figure \ref{figure}) which are the analog of the regions one finds in the 
Penrose diagram of non-extremal spherical black holes \cite{DeLorenzo:2017tgx, DeLorenzo:2018ghq}.

Positive frequency solutions of the massless Klein-Gordon equation, defined with respect to the inertial time $t$,  
can be used  to construct the one-particle Hilbert space $\sH$ of the massless scalar field and then the associated Fock space $\sF$ containing all excitations in Minkowski spacetime. Similarly, there is a natural construction of the Fock space  associated with any of the four regions where the conformal Killing vector field \eqref{kikin} is timelike.  Each arises from the notion the positive frequency solutions with respect to the conformal Killing time in each of these regions. In this paper we will explicitly write the vacuum in $\sF$ in terms of several alternative
expressions in terms of excitations in the other Fock spaces $\sF_{\rm I}$, $\sF_{\rm II}$, $\sF_{\rm III}$ and $\sF_{\rm -III}$.

At the horizon, the conformal Killing field  \eqref{kikin}  satisfies the equation \be \nabla_a(\xi\cdot\xi )\hat =-2\kappa \xi_a, 
\ee  
where
\begin{align}\label{sufy}
	\kappa &\equiv  \frac{2 r_H}{r_0^2 - r_H^2}  
\end{align}
plays the role of the surface gravity in the analogy with black hole \cite{DeLorenzo:2017tgx} and corresponds to the temperature notion $T=\kappa/(2\pi)$ appearing 
in the expression of the Minkowski vacuum that we provide here.

\section{Minkowski modes in spherical coordinates}

In this section we characterise the positive frequency solutions of the massless Klein-Gordon equation with respect to 
inertial time $t$ when written in spherical coordinates. The choice of spherical coordinates is necessary for the decomposition of these modes 
in terms of the modes that are positive frequency with respect with the conformal killing time of the conformal spherical Killing vector fields introduced in what follows. In particular, the spherical coordinates are adapted to the geometry of the lightcones (the conformal Killing horizons) that play a central role in the decomposition of the Minkowski vacuum that we seek.  

We start by writing the Minkowski metric in spherical coordinates
\begin{align}
	\dd s^2 &= - \dd t^2 + \dd r^2 + r^2 \dd \theta^2 + r^2 \sin^2(\theta) \dd \varphi^2 , 
	\end{align}
and the conformally invariant Klein-Gordon equation which, in the previous coordinates and on the flat background, becomes	
\begin{align}
	0 &= \left( \square - \frac{1}{6} R \right)  \Phi (x) \n \\
	&= \frac{1}{\sqrt{-g}} \partial_\mu \left( \sqrt{-g} g^{\mu \nu} \partial_\nu  \right)  \Phi(x) \n \\
	&= \left( -\frac{\partial^2}{\partial t^2} + \frac{1}{r^2} \frac{\partial}{\partial r} \left( r^2 \frac{\partial}{\partial r} \right) + \frac{1}{r^2  \sin(\theta)} \frac{\partial}{\partial \theta} \left(  \sin(\theta) \frac{\partial}{\partial \theta} \right) + \frac{1}{r^2 \sin^2(\theta)} \frac{\partial^2}{\partial \varphi^2} \right)  \Phi (x)
\end{align}
Using the ansatz
\begin{align}
	 \Phi_{\omega \ell m} (x) &= e^{- i \omega t} Y_{\ell m} (\theta, \varphi) R_\ell (r),
\end{align}
the Klein-Gordon equation reads
\begin{align}
	\left( \omega^2 + \frac{\partial^2}{\partial r^2} + \frac{2}{r} \frac{\partial}{\partial r} - \frac{\ell (\ell + 1) }{r^2} \right) R_\ell (r) &= 0, 
\end{align}
with two linearly independent solutions given by the spherical Bessel functions $j_\ell (\omega r)$ and $y_\ell(\omega r)$.  Regularity at the origin discards the $y_\ell(\omega r)$. Thus a basis of the solutions of the Klein-Gordon equation in spherical coordinates is given by
\begin{align}
	 \Phi _{\omega \ell m}(x) &=  \ e^{- i \omega t} \ Y_{\ell m} (\theta, \varphi) \ j_\ell (\omega r) .
\end{align}
Such states are not normalizable in the one-particle Hilbert space $\sH$. Nevertheless it will be convenient to work with them 
formally. Normalizable states peaked on the relevant quantum numbers can be constructed as superpositions of the previous states. 

\subsection{Minkowski-time positivity of frequency as single-ray analyticity}

Solutions of the massless Klein-Gordon equation are completely characterized by their value on the union of the future light cone $u=r_H$ and the past light cone $v=r_H$ with $u\le r_H$. 
In the surface $u=r_H$ 
\begin{align}
	 \Phi_{\omega \ell m}(v,\theta,\varphi) &= Y_{\ell m} (\theta, \varphi) \ e^{- i \omega (v + r_H)/2} j_\ell \left( \frac{\omega (v - r_H )}{2} \right)\ \ \ {\rm for} \ \ \ v\ge r_H,
\end{align}
while for $v=r_H$
\begin{align}
	 \Phi_{\omega \ell m}(u,\theta,\varphi) &= Y_{\ell m} (\theta, \varphi) \ e^{- i \omega (r_H+u)/2} j_\ell \left( \frac{\omega (r_H -u)}{2} \right)\ \ \ {\rm for} \ \ \ u\le r_H.
\end{align}
A generator (light-ray) labelled by $\theta$ and $\varphi$ in the past section of the light cone corresponds to the one labelled by
$\theta=\pi-\theta$ and $\varphi=\varphi+\pi$ in the future section. Under such antipodal map in the sphere one has that 
\be Y_{\ell m} (\pi-\theta, \varphi+\pi)=(-1)^\ell Y_{\ell m} (\theta, \varphi),\ee
which combined with the property of spherical Bessel functions
\be
j_\ell(-x)=(-1)^\ell j_\ell(x)
\ee 
 implies that on a single light-cone generator
 \begin{align}
	 \Phi_{\omega \ell m}(v,\text{ray}) &= A_{\text{ray}} \ e^{- i \omega (v + r_H)/2} j_\ell \left( \frac{\omega (v - r_H )}{2} \right)\ \ \ {\rm for} \ \ \ v\ge r_H
\end{align}
and \begin{align}
	 \Phi_{\omega \ell m}(u,{\rm ray}) &= A_{\rm ray}  \ e^{- i \omega (r_H+u)/2} j_\ell \left( \frac{\omega (u-r_H)}{2} \right)\ \ \ {\rm for} \ \ \ u\le r_H,
\end{align}
where $A_{\rm ray}=Y_{\ell m} (\theta, \varphi)=(-1)^\ell Y_{\ell m} (\pi-\theta, \varphi+\pi)$.
It follows that such a solution, when restricted to a single generator, can be written in terms of a single variable $z \in \mathbb{R}$ which will correspond to either $u$ or $v$ depending on the range. We have
\begin{align}
	\Phi_{\omega \ell m}(z,{\rm ray}) &=  A_{\rm ray} \ e^{- i \omega (z + r_H) / 2} j_\ell \left( \frac{\omega (z - r_H)}{2} \right) .
\end{align}
On a given generator, the previous solutions are given by the product of two entire functions of the variable $z$ \cite{handbook} (now promoted to a complex variable). Thus the previous positive frequency solution corresponds to an analytic function when restricted to a single light-cone generator.  Since the previous function is analytic, it can only diverge at infinity. Now, the asymptotic behaviour of $j_\ell$ at infinity is given by
\begin{align}
	j_\ell (z) &\approx \frac{1}z \sin \left( z - \frac{\ell\pi}{2} \right) .
\end{align}
Hence
\begin{align}\label{eququ}
	 \Phi_{\omega \ell m}(z,{\rm ray}) &\approx \frac{A_{\rm ray}}{\omega \left(z - r_H \right)} e^{-i \omega (r_H + z)/2} \left( e^{i \omega \left(z - r_H \right)/2 + i \ell \pi/2} - e^{-i \omega \left(z - r_H \right)/2- i \ell \pi/2}  \right) \n \\
	&= \frac{A_{\rm ray}}{\omega \left( z - r_H \right)} \left( e^{- i \left(\omega r_H  - \frac{\ell \pi}{2} \right)} - e^{- \frac{i \ell \pi}{2}} e^{- i \omega z} \right) .
\end{align}
From the previous equation we conclude that superpositions of modes with  $\omega>0$ correspond to analytic functions of $z$ that are bounded in the lower complex plane ($\Im(z)<0$). Thus positive frequency solutions of the Klein-Gordon equations are characterised, on the union of the light cones $v=r_H$ and $u=r_H$ when evaluated on a single generator, by analytic functions of $z$ bounded in the lower complex plane.

\begin{figure}[h!!!!!!!!!]
	\centering
	{\includegraphics[scale=0.18]{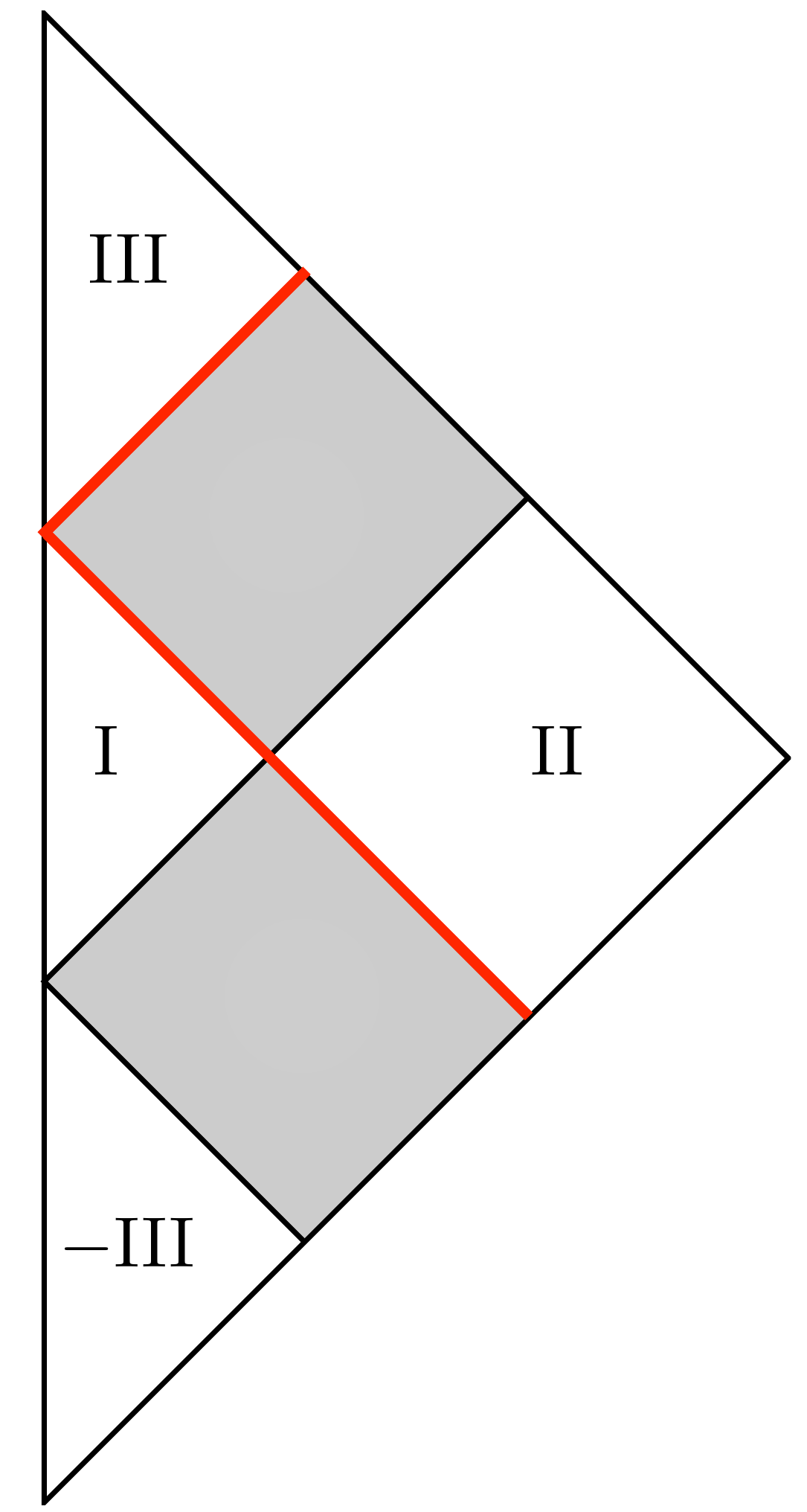}}
	\hspace{10mm}
	{\includegraphics[scale=0.18]{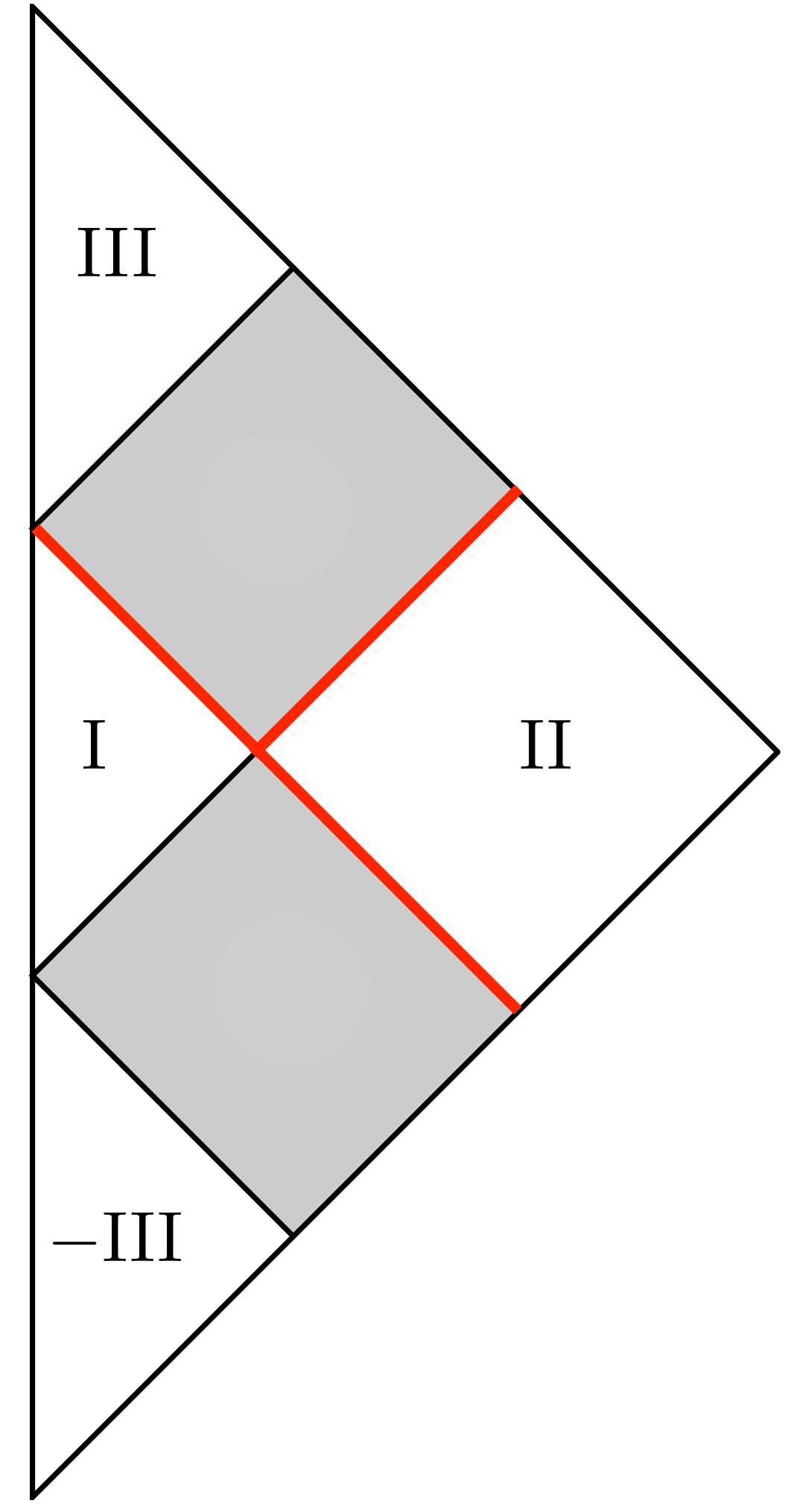}}
	\hspace{10mm}
	{\includegraphics[scale=0.18]{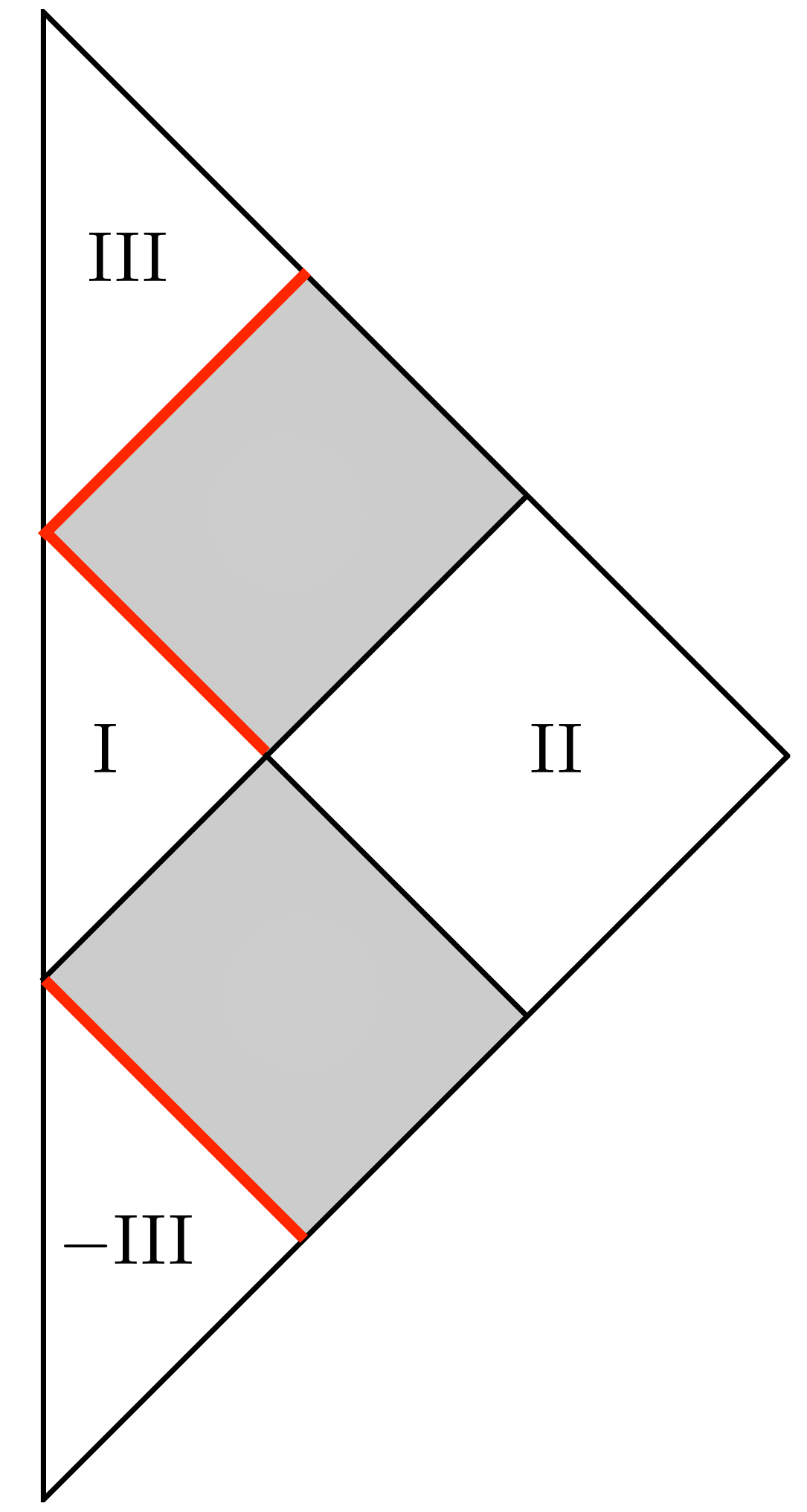}}\hspace{10mm}
	{\includegraphics[scale=0.18]{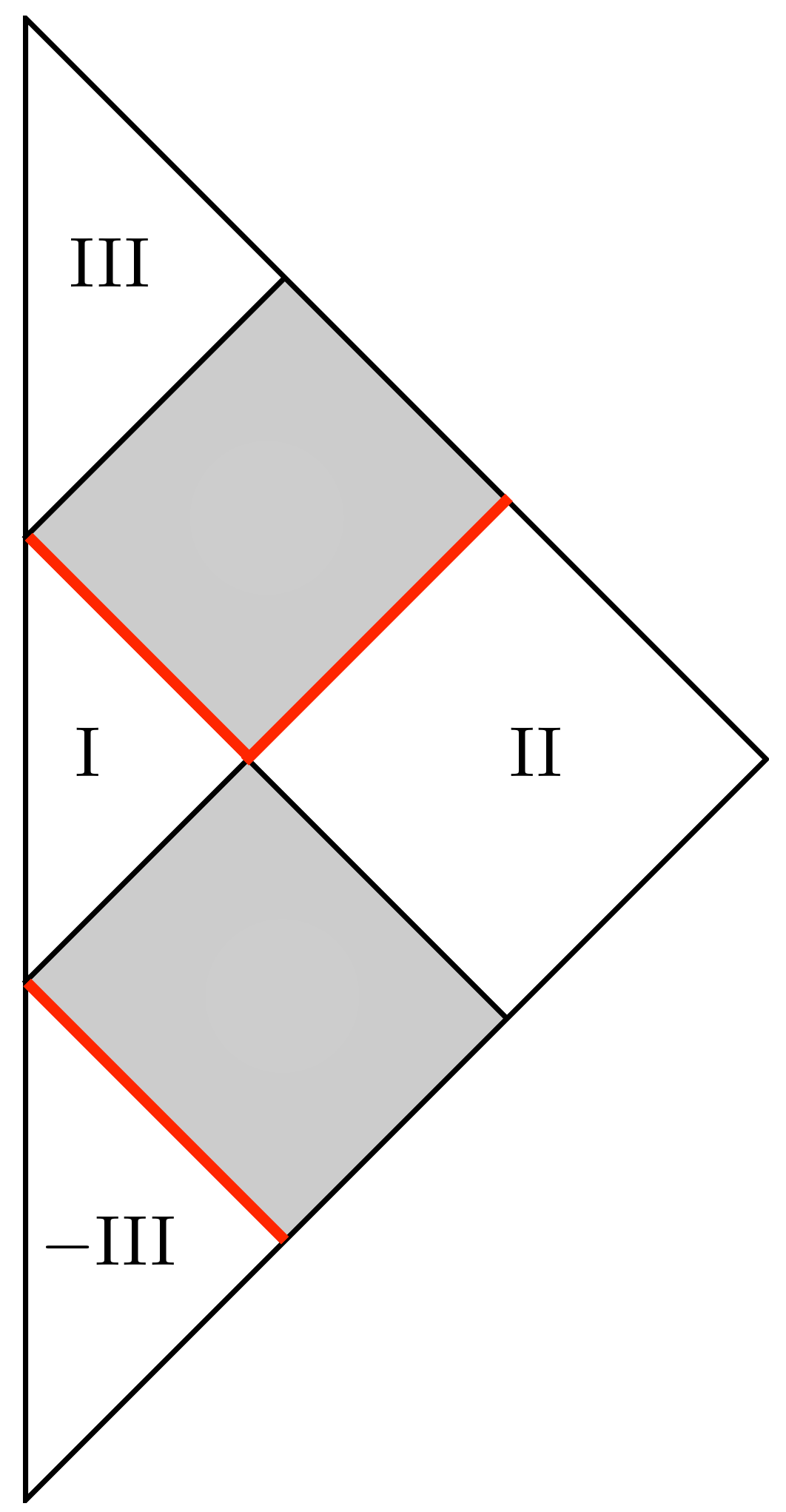}}
	\caption{The causal character of the spherical conformal Minkowski Killing vector field divides flat spacetime in six different regions. The shaded regions correspond to those where the conformal Killing vector field is spacelike. Fock spaces can be constructed according to the positive frequency notion associated with the Killing time in the four white regions where the vector field is timelike. We call such Hilbert spaces $\sF_I$, $\sF_{II}$, $\sF_{III}$, and $\sF_{-III}$ . The Fock space constructed from positive frequency solutions in inertial time will be called $\sF$. Solutions of the massless Klein-Gordon equation in Minkowski spacetime can be fully characterized by their value on the portions of null surfaces emphasized in red. This is the key for the different ways one can express the Minkowski vacuum presented in equations \eqref{UNO},  \eqref{CUATRO}, \eqref{DOS}, and \eqref{TRES}.}
	\label{figure}
\end{figure}

\section{Spherical conformal Killing modes}

In this section we characterize positive frequency solutions with respect to the conformal Killing time
in the different regions where the conformal Killing vector field $\xi^a$ defined in \eqref{kikin} is timelike (see Figure \ref{figure}).
In each of these regions one can define a Fock quantization using standard methods.

\subsection{Regions II, III and -III}

Let us consider the following coordinate transformation \cite{DeLorenzo:2017tgx}
\begin{align}\label{coord}
	t &=  \frac{r_H \sinh(\kappa \tau)}{\cosh(\kappa \rho) - \cosh(\kappa \tau) }, \n \\
	r &= - \frac{r_H \sinh(\kappa \rho)}{\cosh(\kappa \rho) - \cosh(\kappa \tau) }, 
\end{align}	
with $\kappa$ the surface gravity \eqref{sufy}.
In terms of null coordinates the previous transformation takes the form
\begin{align} \label{eq:uv2}
	u &= t - r = - r_H \coth\left( \frac{\kappa \tilde{u}}{2} \right) , \n \\
	v &= t + r = - r_H \coth\left( \frac{\kappa \tilde{v}}{2} \right).
\end{align}
The previous coordinate transformation allows to write the Minkowski metric in regions II, III and -III as  
\begin{align}
	\dd s^2 &= \Omega^2_{\rm II} \left( -\dd \tau^2 + \dd \rho^2 + \kappa^{-2} \sinh^2 (\kappa \rho) \dd S^2 \right) ,
\end{align}
with
\begin{align}
	\label{eq:mink2}
	\Omega_{\RN{2}} &= \frac{r_H \kappa}{ \cosh(\kappa \rho) - \cosh(\kappa \tau) },
\end{align}
where each of the regions is characterized by the range of the coordinates which is best defined in terms of $\tilde u$ and $\tilde v$.
Region II is defined by $\tilde{u} := \tau - \rho \in \R^+$ and $\tilde{v} \equiv \tau + \rho \in \R^-$, Region III is defined by $\tilde{u} := \tau - \rho \in \R^-$ and $\tilde{v} \equiv \tau + \rho \in \R^-$, Region -III is defined by $\tilde{u} := \tau - \rho \in \R^+$ and $\tilde{v} \equiv \tau + \rho \in \R^+$. We would like to characterize solutions of the conformally invariant Klein-Gordon equation
\begin{align}
	\label{eq:CKG}
	\left( \square - \frac{1}{6} R \right) U &= 0
\end{align}
when restricted to (suitable portions of) the boundary of Regions II, III, or -III. Under a conformal transformation $g_{ab} \rightarrow g'_{ab} = C^2 g_{ab}$ solutions of \eqref{eq:CKG} defined in terms of $g_{ab}$ are mapped into solutions of the same equation in terms of $g'_{ab}$ by the rule $\Phi \rightarrow C^{-1} \Phi$ \cite{Wald:1984rg,Birrell:1982ix}. In the new coordinates equation \eqref{eq:CKG} reads
\begin{align}
	 \left[ - \partial_\tau^2 + \frac{1}{\kappa \sinh^2(\kappa \rho)} \left( \partial_\rho \sinh^2(\kappa \rho) \partial_\rho  + \frac{\kappa^2}{\sin(\theta)} \partial_\theta (\sin(\theta) \partial_\theta) + \frac{\kappa^2}{\sin^2(\theta)} \partial^2_\varphi \right) + \kappa^2 \right] U(x)=0
\end{align}
where we used that $R= -6 \kappa^2$. Solutions are given by
\begin{align}
	U_{\omega\ell m} (x) &= e^{- i \omega \tau} \frac{Q^\ell_{\omega\pm}(\rho)}{\sinh(\kappa \rho)} Y_{\ell m} (\theta, \varphi)
\end{align}
with $Q^\ell_{\omega\pm}(\rho)$ satisfying the equation
\begin{align}
	\left( \partial_\rho^2 + \omega^2 - \frac{\ell (\ell + 1) \kappa^2}{\sinh^2 (\kappa \rho)} \right) Q^\ell_{\omega \pm} (\rho) = 0 .
\end{align}
A fact that is central in what follows is that the effective potential $-{\ell (\ell + 1) \kappa^2}/{\sinh^2 (\kappa \rho)}$ vanishes exponentially as one approaches any of the internal null boundaries of Regions II, III, and $-$III so that solutions of the previous equation are well approximated by free waves
\begin{align}
	U_{\omega \ell m} (x) &\approx  \frac{e^{-i \omega (\tau \pm \rho)}}{\sinh(\kappa \rho)} Y_{\ell m} (\theta, \varphi),
\end{align}
which will lead to a simple plane wave functional dependence when restricted to the null boundaries of the corresponding regions.
Explicitly, on the relevant boundaries of the regions II, III, and -III the solutions are either
\begin{align}
	\Phi_{\omega\ell m}  (x) &= \Omega_{\rm II}^{-1} U_{\omega\ell m} = \frac{ e^{-i\omega \tilde{u}}}{r} \ Y_{\ell m}(\theta,\varphi).
\end{align}
or \begin{align}
	\Phi_{\omega\ell m}  (x) &= \Omega_{\rm II}^{-1} U_{\omega\ell m} = \frac{ e^{-i\omega \tilde{v}}}{r} \ Y_{\ell m}(\theta,\varphi),
\end{align}
depending on whether we focus on null boundaries of constant $v$ or $u$ respectively.  In the previous equation we have used \eqref{coord} and \eqref{eq:mink2} to obtain $1/r$ prefactors.
Inverting the relationship \eqref{eq:uv2}
\footnote{Explicitly  we have 
\ba
\tilde u&=&{2}{\kappa^{-1}} \coth^{-1}\left(-\frac{u}{r_H}\right)={\kappa^{-1}} \log\left(\frac{u-r_H}{u+r_H}\right)\n \\ 
\tilde v&=&{2}{\kappa^{-1}} \coth^{-1}\left(-\frac{v}{r_H}\right)={\kappa^{-1}} \log\left(\frac{v-r_H}{v+r_H}\right).\n 
\ea
}
 we can express the solutions  with a definite frequency $\omega$---as defined by the accelerated conformal observers---on the past boundary $v=r_H$ of Region II (its past horizon) as 
\begin{align}\label{uno}
	\Phi^{\RN{2}}_{\omega \ell m}(x) &= \frac{1}{r_H - u} Y_{\ell m} (\theta, \varphi) \ e^{- i \frac{\omega}{\kappa} \log\left( \frac{u - r_H}{u + r_H} \right)  },
\end{align}
for $u\le -r_H$.
While for the future horizon $u=-r_{H}$ of Region II 
\begin{align}\label{dos}
	\Phi^{\rm II}_{\omega \ell m}(x) &= \frac{1}{v+ r_H} Y_{\ell m} (\theta, \varphi) \ e^{- i \frac{\omega}{\kappa} \log\left( \frac{v - r_H}{v + r_H} \right)  },
\end{align}
for $v\ge r_H$.
Similarly, for the boundary of Region $-$III, $v=-r_H$, we have
\begin{align}\label{unos}
	\Phi^{-\rm III}_{\omega \ell m}(x) &= \frac{1}{u + r_H} Y_{\ell m} (\theta, \varphi) \ e^{- i \frac{\omega}{\kappa} \log\left( \frac{u - r_H}{u + r_H} \right)  },
\end{align}
with $u\le r_H$, while for the boundary of Region III, $u=r_H$, we get
\begin{align}\label{doss}
	\Phi^{\rm III}_{\omega \ell m}(x) &= \frac{1}{v - r_H} Y_{\ell m} (\theta, \varphi) \ e^{- i \frac{\omega}{\kappa} \log\left( \frac{v - r_H}{v + r_H} \right)  },
\end{align}
with $v\ge r_H$. Note that the functional form of the modes on the null boundaries of interest is always the same while the range and nature of the variables is different. 

\subsection{Region I}

Modes in Region I can be described in a way similar to what we have done in the previous section. Instead of \eqref{coord} one needs to
consider the coordinate transformation
\begin{align}\label{coord1}
	t &= \frac{r_H \sinh(\kappa \tau )}{\cosh(\kappa \rho) + \cosh(\kappa \tau) } , \n \\
	r &= \frac{r_H \sinh(\kappa \rho)}{\cosh(\kappa \rho) + \cosh(\kappa \tau) } .
\end{align} 
With the new transformation the relation \eqref{eq:uv2} is replaced by
\begin{align}\label{laba}
	v &= r_H \tanh \left( \frac{\kappa \tilde{v}}{2} \right) ,\n  \\
	u &= r_H \tanh \left( \frac{\kappa \tilde{u}}{2} \right).
	\end{align}
For $\tilde u\in \R$ and $\tilde v\in \R$ the Minkowski metric
in Region I reads	
\begin{align}
\DIFaddbegin \label{eq:MinkI}
	\DIFaddend \dd s^2 &= \Omega_{\rm I}^2 \left( - \dd \tau^2 + \dd \rho^2 + \kappa^{-2} \sinh^2 (\kappa \rho) \dd S^2 \right), 
\end{align}
where
\begin{align}
	\Omega_{\rm I} &= \frac{r_H \kappa}{\cosh(\kappa \rho) + \cosh(\kappa \tau)} .
\end{align}
It follows from the same arguments that the solutions of equation \eqref{eq:CKG} on the future null boundary of Region I,  $v=r_H$, are given by
\begin{align}
	\Phi^I_{\omega\ell m}(x) &= \frac{1}{r} Y_{\ell m} (\theta, \varphi) e^{- i \omega \tilde{u}}.
\end{align}
Using \eqref{laba} we can express the modes as a function of Minkowski retarded time
\footnote{The inverse transformation being in this case 
\ba
\tilde u&=&{2}{\kappa^{-1}} \tanh^{-1}\left(\frac{u}{r_H}\right)={\kappa^{-1}} \log\left(\frac{u+r_H}{r_H-u}\right)\n \\ 
\tilde v&=&{2}{\kappa^{-1}} \tanh^{-1}\left(\frac{v}{r_H}\right)={\kappa^{-1}} \log\left(\frac{v+r_H}{r_H-v}\right).\n 
\ea
}
\begin{align}\label{tres}
	\Phi^{\RN{1}}_{\omega \ell m}(x) &= \frac{1}{r_H - u} Y_{\ell m} (\theta, \varphi) e^{- \frac{i \omega}{\kappa} \log\left( \frac{r_H + u}{r_H - u} \right) }.
\end{align}
Similarly, on the past null boundary of Region I, $u=-r_{H}$, the modes are
\begin{align}\label{cuatro}
	\Phi^{\RN{1}}_{\omega \ell m}(x) &= \frac{1}{ v + r_H } Y_{\ell m} (\theta, \varphi) e^{- \frac{i \omega}{\kappa} \log\left( \frac{r_H + v}{r_H - v} \right)}.
\end{align}
All these solutions diverge at $r = 0$ due to the vanishing of the conformal factor at that singular point. 
This is a pathology of the sharp `plane-wave-like' solutions; however, such a divergence cannot survive if we consider suitably 
normalized wave packets satisfying the usual reflecting boundary conditions at $r=0$.

\section{Purification}

The knowledge of the modes on the null boundaries 
of the different regions I, II, III, and $-$III in Figure \ref{figure}, as well as their characterization in terms of frequencies with respect to the conformal Killing time $\tau$, are sufficient for writing an explicit expression of the Minkowski vacuum in terms of particle excitations in the Fock quantizations corresponding to these various regions.  

\subsection{Vacuum entanglement between regions I, outgoing II and III}

In order to find the expression of the Minkowski vacuum in terms of the product of states in $\sF_{\rm I}\otimes\left(  \sF_{\rm II}^{\rm out} \oplus \sF_{\rm III} \right)$ associated to the
Regions I, II (for outgoing modes only), and III, respectively, we focus on the form of the solution of the Klein-Gordon equation on the red null boundaries on the panel on the left of Figure \ref{figure}.  It is clear that the value of solutions on these null surfaces fully determine the solution everywhere. We can translate this statement in terms of the relevant one-particle Hilbert spaces involved in what follows. The one-particle Hilbert space $\sH_{\rm I}$  is completely characterized by the value of the (normalizable) positive frequency solutions with respect to the conformal killing time in Region I when restricted to the future null boundary of Region I.
The one-particle Hilbert space  $\sH_{\rm II}^{\rm out}$ of outgoing modes in region II is completely characterized by the value of the (normalizable) positive frequency solutions with respect to the conformal killing time in Region II when restricted to the past null boundary of Region II. The one-particle Hilbert space $\sH_{\rm III}$ is completely characterized by the value of the (normalizable) positive frequency solutions with respect to the conformal Killing field in Region III,  when restricted to the past null boundary of Region III. The one-particle Hilbert space of positive frequency solutions with respect to inertial time $t$ will be denoted by $\sH$.

Now, we will construct a Minkowski inertial time positive frequency solution by combining definite (conformal time) frequency solution in Regions I, II, and III. 
According to our previous analysis---recall equations \eqref{uno}, \eqref{dos},  and \eqref{tres}---these are given by 
\begin{align}\label{pitito}
	f_{\omega \ell m}^{\rm IIout} &= \frac{(-1)^\ell Y_{\ell m}(\theta, \varphi)}{r_H - u} \exp \left(- \frac{i \omega}{\kappa} \log \left( \frac{u- r_H}{u + r_H} \right) \right) , \quad u < - r_H \ \ \ \ {\rm (the\ past\ horizon\ of\ II)} \n \\
	f_{\omega \ell m}^{\RN{1}} &= \frac{(-1)^\ell Y_{\ell m}(\theta, \varphi)}{r_H - u} \exp\left( \frac{i \omega}{\kappa} \log \left( \frac{r_H - u}{u + r_H} \right) \right) , \quad -r_H\le u \le r_H , \ \ \ \ {\rm (the \ future \ horizon\ of \ I)} \n \\
        f_{\omega \ell m}^{\RN{3}} &= \frac{Y_{\ell m}(\theta, \varphi)}{r_H - v} \exp\left( - \frac{i \omega}{\kappa} \log \left( \frac{ v - r_H }{v + r_H} \right) \right), \quad\ \ \ \ \ \ \ \, v  > r_H , \ \ \ \ {\rm (the \ past \ horizon\ of \ III)} ,
\end{align}
where the $(-1)^\ell$ has been included as to simplify the expressions that follow when writing  a positive frequency solution in inertial Minkowski time (recall the need for single-ray analyticity). 
Consequently, focusing on a single  generator of the light cone and
using the variable $z$ to represent both $u$ and $v$, we can write  
\begin{align}
	f_\omega^{\rm IIout/III} &= \frac{1}{r_H - z} \exp\left( - \frac{i \omega}{\kappa} \log \left( \frac{z - r_H}{z + r_H} \right) \right), \quad \left| z \right| > r_H , \n \\
	f_\omega^{\RN{1}} &= \frac{1}{r_H - z} \exp\left( \frac{i \omega}{\kappa} \log \left( \frac{r_H - z}{z + r_H} \right) \right) , \quad \ \  \, \left| z \right| < r_H .
\end{align}
For $\omega>0$, these modes are positive frequency with respect to the time notion associated with the conformal `observers' time $\tau$ defined in regions II/III and I, respectively. 
Let us promote $z$ to a complex variable and consider  the function
\begin{align}
	F_\omega (z) &= \frac{1}{r_H - z} \exp \left( - \frac{i \omega}{\kappa} \log \left( \frac{z - r_H}{z + r_H} \right) \right) .
\end{align}
This function is analytic everywhere with the exception of the real interval $\left[ - r_H , r_H \right]$ where it has a branch cut and a pole at $z= r_H$ (which corresponds to $r=0$).
The branch cut is present due to the infinite blue shift effect approaching the null boundaries for the conformal observers with conformal Killing time. The pole at $z= r_H$ is due to the vanishing of the conformal factor at $r=0$; this pole is present in the wave solutions that we use here for simplicity and it  disappears when considering a suitable basis of normalizable wave packets with the customary reflecting boundary conditions at the origin.  Hence, restricting to lower complex plane, the previous function is analytic and bounded; therefore, according to the analysis below equation \eqref{eququ}, it is a positive-frequency mode with respect to inertial time time $t$. Indeed this function can be seen as the restriction of a positive frequency mode on the same components of null boundaries where equation \eqref{pitito} is evaluated, and hence fully determining a unique solution of the Klein-Gordon equation.

 Evaluating the previous solution near the real line from below, namely at $z=u - i \epsilon$, with $u$ real and $\epsilon>0$, we see that it coincides with  $f_\omega^{\rm II/III}$ in the limit $\epsilon \rightarrow 0$ for $|u|>r_{H}$. Now for $-r_H < u < r_H$ the situation is more subtle due to the presence of the branch cut when $\epsilon=0$. 
Indeed we have
\begin{align}
	F_\omega \left( u - i \epsilon \right) &= \frac{1}{r_H - u + i \epsilon} \exp \left( - \frac{i \omega}{\kappa} \log \left( \frac{u - r_H - i \epsilon}{u + r_H - i \epsilon} \right) \right) \n \\
	&= \frac{1}{r_H - u + i \epsilon} \exp \left( - \frac{i \omega}{\kappa} \log \left( \frac{r_H - u - i \epsilon}{u + r_H - i \epsilon} e^{- i (\pi - \mathcal{O} (\epsilon) ) } \right) \right) \n \\
	&= \frac{1}{r_H - u + i \epsilon} e^{- \frac{(\pi - \mathcal{O} (\epsilon) ) \omega}{\kappa}} \exp \left( - \frac{i \omega}{\kappa} \log \left( \frac{r_H - u - i \epsilon}{u + r_H - i \epsilon} \right) \right) \n \\
	&\xrightarrow[\epsilon \rightarrow 0]{} e^{- \frac{\pi \omega}{\kappa}}  \overline{f}_\omega^{\ \RN{1}} .
\end{align}
Thus we have showed that 
\begin{align}
	F_\omega (u) &= f_\omega^{\rm IIout} + f_\omega^{\rm III} + e^{- \frac{\pi \omega}{\kappa}}  \overline{f}_\omega^{\RN{1}}
\end{align}
is positive frequency in Minkowski time. If instead of working with the previous non-normalizable states we build a basis of wave packets $f_\omega$ sharply peaked about the frequency $\omega$, then the following combination is also positive frequency in Minkowski time:
\begin{align}
	F_\omega (u) &= f_\omega^{\rm II-out} + f_\omega^{\rm III} + e^{- \frac{\pi \omega}{\kappa}}  \overline{f}_\omega^{\RN{1}}.
\end{align}
We can repeat the procedure considering the function
\begin{align}
	F'_\omega (z) &=  \frac{1}{r_H - z} \exp \left(  \frac{i \omega}{\kappa} \log\left( \frac{r_H - z}{r_H + z} \right) \right) .
\end{align}
This function is analytic everywhere but in the real intervals $\left(-\infty, -r_H \right]$ and $\left[ r_H, +\infty \right)$ and at the pole $z = r_H$. As before, we evaluate $F'_\omega$ at $z = u - i \epsilon$, with $u$ real and $\epsilon > 0$. For $- r_H < u < r_H$ it gives $f_\omega^{{\rm I}}$ in the limit $\epsilon \rightarrow 0$. For $u> r_H$ we have
\begin{align}
	F'_\omega \left(u - i\epsilon \right) &= \frac{1}{r_H - u + i \epsilon}  \exp \left( \frac{i\omega}{\kappa} \log \left( \frac{r_H - u + i \epsilon}{r_H + u - i \epsilon} \right) \right) \n \\
	&= \frac{1}{r_H - u + i \epsilon} \exp \left( \frac{i\omega}{\kappa} \log \left( \frac{u - r_H + i\epsilon}{u + r_H - i \epsilon} e^{i (\pi - \mathcal{O} (\epsilon) ) } \right) \right) \n \\
	&= \frac{1}{r_H - u + i \epsilon} e^{- \frac{(\pi - \mathcal{O}(\epsilon) ) \omega}{\kappa}} \exp \left( \frac{i\omega}{\kappa} \log \left( \frac{u - r_H + i\epsilon}{u + r_H - i \epsilon} \right) \right)\n  \\
	&\xrightarrow[\epsilon \rightarrow 0]{} e^{-\frac{\pi \omega}{\kappa}}  \overline{f}_\omega^{\RN{3}} .
\end{align}
The same result applies for $u<-r_H$ (with $\overline{f}_\omega^{\rm II-out}$ instead of $\overline{f}_\omega^{\rm III}$). Thus also the following linear combination is positive frequency
\begin{align}
	F'_\omega (u) &= f_\omega^{\RN{1}} + e^{-\frac{\pi \omega}{\kappa}}  \left( \overline{f}_\omega^{\rm IIout} + \overline{f}_\omega^{\rm III} \right),
\end{align}
with the wave packet argument being still valid. Hence we can compute the S-matrix $S$. By inspection
\begin{align}
	C F_\omega = f_\omega^{\rm II-out} + f_\omega^{\rm III} &, \qquad C F_\omega' = f_\omega^\RN{1}\n \\
	D F_\omega = e^{- \frac{\pi \omega}{\kappa}}  \overline{f}_\omega^{\ \RN{1}} &, \qquad D F_\omega' = e^{-\frac{\pi \omega}{\kappa}} \left(  \overline{f}_\omega^{\rm II-out} + \overline{f}_\omega^{\rm III} \right) ,
\end{align}
where $C: \sH\to \sH_{\rm I}\otimes \left(\sH_{\rm IIout} \oplus \sH_{\rm III} \right)$ is the map that gives the positive conformal frequency part of an inertial time positive frequency solution, and
$D:  \sH\to \overline \sH_{\rm I}\otimes \left( \overline \sH_{\rm IIout} \oplus  \overline \sH_{\rm III}  \right) $ is the map that gives the negative conformal frequency part of an inertial time positive frequency solution.  It follows from the previous equations that
\begin{align}
	D C^{-1} \left( f_\omega^{\rm II-out} + f_\omega^{\rm III} \right) &= e^{-\frac{\pi \omega}{\kappa}}  \overline{f}_\omega^{\ \RN{1}} , \n \\
	D C^{-1} f_\omega^{\RN{1}} &= e^{-\frac{\pi \omega}{\kappa}} \left( \overline{f}_\omega^{ \rm IIout} + \overline{f}_\omega^{\rm III}  \right).
\end{align}
Since $\left\{ f_\omega^\RN{1}  \right\}$ and  $\left\{ f_\omega^{\rm II-out} + f_\omega^{\rm III}  \right\}$ jointly span $\mathcal{\sH}_2 = \mathcal{\sH}_\RN{1} \otimes \left( \sH_{\rm II-out} \oplus \sH_{\rm III} \right)$, the previous equation determines the fundamental two-particle state $\mathcal{E} = \overline{D} \overline{C}^{-1}$ which can be written as 
\begin{align}
	\mathcal{E}^{ab} &= 2 \sum_\omega e^{- \frac{\pi \omega}{\kappa}} f_{\omega \RN{1}}^{(a} \ \left( f_{\omega\rm IIout} + f_{\omega \rm III} \right)^{b)}.
\end{align}
A standard construction \cite{Wald:1984rg} leads to the expression of the state $S\ket{0}$ (where $\ket{0}$ denotes the Minkowski vacuum), namely
\be\boxed{S \left| 0 \right>_M = \prod_{\omega \ell m} \left( \sum_{n=0}^{\infty} e^{- \frac{n \pi \omega }{\kappa}} \left| n, \omega,\ell,m \right>_{\RN{1}} \otimes \left( \left| n, \omega, \ell, m \right>_{ {\rm II}-{\rm out}} \oplus  \left| n, \omega, \ell, m \right>_{ \rm III} \right) \right) }, \label{UNO}
\ee
where $\left| n, \omega \right>$ are Fock space `basis' states with a definite number of particles $n$ in the conformal frequency mode $\omega_i$ in the regions I, the outgoing modes of region II, or all the modes of region III, respectively. This is our main result. It clearly shows that the reduced density matrix obtained from tracing $\ket{0}\bra{0}$ over $\sH_{\rm I}$ is thermal with temperature $T=\kappa/(2\pi)$ as in \cite{DeLorenzo:2017tgx}.

\subsection{Other equivalent purifications}

It is possible to rewrite equation \eqref{UNO} in different equivalent forms. 
For instance from the form of the conformal invariant Klein Gordon equation in the 
region where the conformal Killing field is spacelike (the grey regions in Figure \eqref{figure}) it is 
easy to see that an ingoing positive frequency solution in Region II corresponds to a positive frequency solution in Region III (see Appendix \ref{app}).   
This allows for a trivial identification between the Hilbert spaces $\sF_{\rm III}$ and $\sF_{\rm II-in}$, and the rewriting of equation \eqref{UNO} as
\be\label{CUATRO}
\boxed{	U \left| 0 \right>_M = \prod_{\omega \ell m} \left( \sum_{n=0}^{\infty} e^{- \frac{n\pi\omega}{\kappa}} \left| n, \omega, \ell, m\right>_\RN{1} \otimes \left| n, \omega, \ell, m \right>_{{\rm II}}  \right) }.
\ee
Similarly, one can write using the trivial isomorphism between $\sF_{\rm -III}$ and $\sF_{\rm II-out}$
\be\label{DOS}
\boxed{	U \left| 0 \right>_M = \prod_{\omega \ell m} \left( \sum_{n=0}^{\infty} e^{- \frac{n\pi\omega}{\kappa}} \left| n, \omega, \ell, m \right>_\RN{1} \otimes \left( \left| n, \omega, \ell, m \right>_{-\RN{3}} \oplus \left| n, \omega, \ell, m \right>_{\RN{3}} \right) \right) }.
\ee
Finally,  the time reverse of \eqref{UNO} also holds
%
\be\label{TRES}
\boxed{	U \left| 0 \right>_M = \prod_{\omega \ell m} \left( \sum_{n=0}^{\infty} e^{- \frac{n\pi\omega}{\kappa}} \left| n, \omega, \ell, m \right>_\RN{1} \otimes \left( \left| n, \omega, \ell, m \right>_{-\RN{3}} \oplus \left| n, \omega, \ell, m \right>_{\rm II-in} \right) \right) }.
\ee
In the extremal case $\kappa=0$, where Region I shrinks to a point, one can show that the Minkoswki vacuum coincides with the vacuum of the
conformal observers associated to \eqref{kikin}. The reason can be traced to the absence of a branch cut in the relationship between the two notions of retarded and advanced times (see appendix in \cite{DeLorenzo:2017tgx}). 
\DIFaddbegin 

\section{{Conformal Killing observers are uniformly accelerating}} 

{An observer following one integral line of the conformal Killing field has four velocity given by
}\begin{align}
	{u^a = \frac{\xi^a}{\sqrt{- \xi_b \xi^b}} = \sqrt{\frac{v^2 - r_H^2}{u^2 - r_H^2}} \left(\frac{\partial}{\partial v} \right)^a + \sqrt{\frac{u^2 - r_H^2}{v^2 - r_H^2}} \left(\frac{\partial}{\partial u} \right)^a .
}\end{align}
{This four velocity describes radially accelerating observers with constant acceleration $a^b := u^a \nabla_a u^b$, i.e., ${u^a \nabla_a \left(a \cdot a\right) }= 0 $. The magnitude of the acceleration is explicitly given by
}\begin{align}
	{\left| a \right| = \sqrt{a_\mu a^\mu} }&{=\kappa \frac{ \ r}{r_H} \frac{1}{\sqrt{\xi_\mu \xi^\mu}}
}\end{align}
Standard results in quantum field theory imply that such uniformly accelerating observer will sense a temperature $T_{\rm obs}=|a|/(2\pi)$; for instance if sensed by an idealized Unruh-DeWitt detector following such orbits.  The 
temperature measured differs from the conformal invariant notion $T\equiv \kappa/(2\pi)$---appearing in the 
purification our formulae---by the constant quantity   ${r}/({r_H}\sqrt{\xi_\mu \xi^\mu})$. The mismatch can be understood as follows: 
the temperature in our purification formulae refers to a conformally invariant property of the Minkowski vacuum of a conformally invariant scalar field. Such notion coincides with a physical temperature
only in the conformal geometry where $\xi$ is an actual Killing field \cite{DeLorenzo:2018ghq}. In all other conformally related geometries, physical thermometers break conformal invariance and 
additional geometric considerations are needed in order to relate 
 $T$ and $T_{\rm obs}$. One can be explicit in the case of the Minkowski metric written as in Eq.\eqref{eq:MinkI} whose Euclidean continuation obtained via the replacement $\tau\to -i\tau_E$ is
\begin{align}
	{\dd s^2 }&{= \Omega_{\rm I}^2\left(-i \tau_E, \rho \right) \left( \dd \tau_E^2 + \dd \rho^2 + \kappa^{-2} \sinh^2 (\kappa \rho) \dd S^2 \right), }
\end{align}
with
\begin{align}
	{\Omega_{\rm I} (-i \tau_E, \rho) }&{= \frac{r_H \kappa}{\cosh(\kappa \rho) + \cos(\kappa \tau_E)} .
}\end{align}
{The conformal invariant termality of the Minkowski vacuum---at the root of our purification formulae---resides in the periodicity of the metric with period $2\pi / \kappa$ in imaginary conformal Killing time $\tau_E$.
Regularity of the conformal transformations imply that such periodicity cannot be changed hence the conformal invariant character of $T\equiv \kappa/(2\pi)$. Now the physical temperature $T_{\rm obs}$ measured by a local 
Unruh-DeWitt devise is sensitive to the geometric periodicity in imaginary proper time. Such period depends on the conformal factor and can be computed as follows 
}\begin{align}
	{\oint_\tau \dd s }&{= \int_0^{\frac{2 \pi}{\kappa}} \Omega_{\rm I} \dd \tau_E 
	= \frac{2\pi}{ \kappa} \left(\frac{r_H }{ r}\sqrt{\xi_\mu \xi^\mu}\right) = \frac{2\pi}{\left| a \right|}.
}\end{align}
Thus, even when a single observer following one single orbit of the conformal killing field coincides with an Unruh observer, the form of the Minkowski 
vacuum state for a conformally invariant scalar field depends (as usual in quantum field theory) on non-local features which preclude the naive comparison 
when using point-like probes.

\section{Discussion}

We have explicitly written the Minkowski vacuum in terms of the particle modes defined by observers moving along spherical conformal killing 
vector fields. These observers represent accelerated observers moving radially away from a sphere of radius $r_H$ and have causal horizons (conformal Killing horizons) which are given by the light surfaces emanating from that sphere at $t=0$. The formula we derive is the analog of the one derived by Unruh
in terms of Rindler particle states associated with constantly accelerated observers following the boost Killing vector field.  We have found the result by exploiting the analyticity properties that define positive frequency solutions in inertial time. A direct derivation using Bogoliubov coefficients computed via the suitably defined Klein-Gordon inner product may be available but does not seem as the most direct avenue to the final expressions (the use of the characterization of modes in terms of null surfaces is natural and simple using our techniques).  A conformal transformation maps Minkowski spacetime to a portion of the Bertotti-Robinson spacetime \cite{DeLorenzo:2018ghq} which describes in a suitable approximation near-horizon physics of a near-extremal black hole. It is potentially interesting to consider using our result to analyse features of quantum field theory on such backgrounds \cite{Ottewill:2012mq}. The massless scalar field quantum theory has been chosen for simplicity. We expect that  our results  should  naturally generalize to any (free) conformal invariant model of quantum fields.  

 \section{Acknowledgement}

We thank Stefano Galanda, Antony Speranza, and Daniel Sudarsky for useful discussions. We would like to thank the hospitality of the {\em John Bell Institute} and the discussions and inputs we received from participants of the workshop on the {\em Black Hole Information Puzzle} in 2022. This publication was made possible through the support of the ID\# 62312 grant from the John Templeton Foundation, as part of the \href{https://www.templeton.org/grant/the-quantuminformation-structure-ofspacetime-qiss-second-phase}{`The Quantum Information Structure of Spacetime' Project (QISS)}. The opinions expressed in this project/publication are those of the author(s) and do not necessarily reflect the views of the John Templeton Foundation.

\begin{appendix}

\section{Solutions of the Klein-Gordon equations adapted to the conformal Killing field in the regions where it is spacelike}\label{app}

In this section we briefly describe some key features of the solutions of the Klein Gordon equation in the grey regions of Figure \ref{figure} where the conformal Killing field is spacelike.
For concreteness we focus on the region to the future of Region II which we call Region V (the analysis is basically the same in the other one). The coordinate transformation of interest is given by
\begin{align}
	t &= \frac{r_H \cosh\left( \kappa \tau \right)}{\sinh(\kappa \tau) + \sinh(\kappa \rho) }, \\
	r &= \frac{r_H \cosh\left( \kappa \rho \right)}{\sinh(\kappa \tau) + \sinh(\kappa \rho) } ,
\end{align}
with $\tau, \rho \in \mathbb{R}^+$. The double null coordinates are given by
\begin{align}
	v &= t + r = r_H \coth \left( \frac{\kappa \tilde{v}}{2} \right) , \\
	u &= t - r = r_H \tanh \left( \frac{\kappa \tilde{u}}{2} \right) ,
\end{align}
with $u \in \R$ while $v \in \R^+$. In these coordinates, which only cover Region V, the Minkowski metric reads
\begin{align}
	\dd s^2 &= \Omega_{\rm V}^2 \left( \dd \tau^2 - \dd \rho^2 + \kappa^{-2} \cosh^2 \left( \kappa \rho \right) \dd S^2 \right) , 
\end{align}
with conformal factor given by
\begin{align}
	\Omega_{\rm V} &= \frac{ r_H \kappa}{\sinh(\kappa \tau) + \sinh(\kappa \rho)} .
\end{align}
The Klein-Gordon equation, $\left( \square - \frac{R}{6} \right) U = 0$, in the conformal metric $ds^2/\Omega_{\rm V} ^2$ reads
\begin{align}
	  \left( - \frac{1}{\cosh^2 (\kappa \rho)} \partial_\rho \left( \cosh^2(\kappa \rho) \partial_\rho \right) + \partial_\tau^2 +   \frac{\kappa^2}{\cosh^2(\kappa \rho)}\left( \frac{1}{\sin \theta} \partial_\theta \left( \sin\theta \partial_\theta \right) + \frac{1}{\sin^2 \theta} \partial_\varphi^2 \right)-\kappa^2 \right) U=0.
\end{align}
With the ansatz
\be
	U_{\omega\ell m} = e^{-i \omega \tau} \frac{Q_{\omega \ell}(\rho)}{\cosh(\kappa \rho)} Y_{\ell m} (\theta, \varphi) ,\ee
the Klein-Gordon equation reduces to
\be	\left( \omega^2 + \frac{\partial^2}{\partial \rho^2} + \frac{\ell (\ell +1) \kappa^2}{\cosh^2(\kappa \rho)} \right) Q_{\omega \ell} (\rho) = 0 ,
\ee
which, on the boundary, reduces to the same wave equation one finds in all the other regions.
The consequence of this is that there is a one-to-one correspondence between positive frequency solutions in Region III and 
in-going positive frequency solutions in Region II. The same holds true for solutions in region $-$III and out-going solutions in Region II.
The reason is that the quantum number $\omega$ is conserved across the boundaries as the boundary characteristic data 
coincide. This implies that one can have a trivial identification between wave packets defining a basis of the one-particle Hilbert spaces $\sH_{\rm III}$ and $\sH^{\rm in}_{\rm II}$, as well as
between element of a basis of $\sH_{\rm -III}$ and $\sH^{\rm out}_{\rm II}$.

\end{appendix}

\providecommand{\href}[2]{#2}\begingroup\raggedright\endgroup


\begin{thebibliography}{10}

\bibitem{DeLorenzo:2017tgx}
T.~De~Lorenzo and A.~Perez, ``{Light Cone Thermodynamics},'' Phys. Rev. D {\bf
  97} (2018), no.~4, 044052, \href{http://arXiv.org/abs/1707.00479}{{\tt
  arXiv:1707.00479}}.

\bibitem{DeLorenzo:2018ghq}
T.~De~Lorenzo and A.~Perez, ``{Light Cone Black Holes},'' Phys. Rev. D {\bf 99}
  (2019), no.~6, 065009, \href{http://arXiv.org/abs/1811.03667}{{\tt
  arXiv:1811.03667}}.

\bibitem{Kay:1988mu}
B.~S. Kay and R.~M. Wald, ``{Theorems on the Uniqueness and Thermal Properties
  of Stationary, Nonsingular, Quasifree States on Space-Times with a Bifurcate
  Killing Horizon},'' Phys. Rept. {\bf 207} (1991) 49--136.

\bibitem{Hislop:1981uh}
P.~D. Hislop and R.~Longo, ``{Modular Structure of the Local Algebras
  Associated With the Free Massless Scalar Field Theory},'' Commun. Math. Phys.
  {\bf 84} (1982) 71.

\bibitem{Martinetti:2002sz}
P.~Martinetti and C.~Rovelli, ``{Diamonds's temperature: Unruh effect for
  bounded trajectories and thermal time hypothesis},'' Class. Quant. Grav. {\bf
  20} (2003) 4919--4932, \href{http://arXiv.org/abs/gr-qc/0212074}{{\tt
  arXiv:gr-qc/0212074}}.

\bibitem{Martinetti:2008ja}
P.~Martinetti, ``{Conformal mapping of Unruh temperature},'' Mod. Phys. Lett. A
  {\bf 24} (2009) 1473--1483, \href{http://arXiv.org/abs/0803.1538}{{\tt
  arXiv:0803.1538}}.

\bibitem{Casini:2011kv}
H.~Casini, M.~Huerta, and R.~C. Myers, ``{Towards a derivation of holographic
  entanglement entropy},'' JHEP {\bf 05} (2011) 036,
  \href{http://arXiv.org/abs/1102.0440}{{\tt arXiv:1102.0440}}.

\bibitem{Jacobson:2022gmo}
T.~Jacobson and M.~R. Visser, ``{Entropy of causal diamond ensembles},''
  \href{http://arXiv.org/abs/2212.10608}{{\tt arXiv:2212.10608}}.

\bibitem{Jacobson:2018ahi}
T.~Jacobson and M.~Visser, ``{Gravitational Thermodynamics of Causal Diamonds
  in (A)dS},'' SciPost Phys. {\bf 7} (2019), no.~6, 079,
  \href{http://arXiv.org/abs/1812.01596}{{\tt arXiv:1812.01596}}.

\bibitem{Unruh:1976db}
W.~Unruh, ``{Notes on black hole evaporation},'' Phys.Rev. {\bf D14} (1976)
870.

\bibitem{handbook}
F.~Olver, D.~Lozier, R.~Boisvert, and C.~Clark, {\em NIST Handbook of
  Mathematical Functions}, p.~262.
\newblock Cambridge University Press, 01, 2010.

\bibitem{Wald:1984rg}
R.~Wald, {\em General Relativity}.
\newblock University of Chicago Press, Chicago, 1984.

\bibitem{Birrell:1982ix}
N.~D. Birrell and P.~C.~W. Davies, {\em {Quantum Fields in Curved Space}}.
\newblock Cambridge Monographs on Mathematical Physics. Cambridge Univ. Press,
  Cambridge, UK,
1984.
\newblock

\bibitem{Ottewill:2012mq}
A.~C. Ottewill and P.~Taylor, ``{Quantum field theory on the Bertotti-Robinson
  space-time},'' Phys. Rev. D {\bf 86} (2012) 104067,
  \href{http://arXiv.org/abs/1209.6080}{{\tt arXiv:1209.6080}}.

\end{thebibliography}
\end{document}